\documentclass[preprint,12pt]{elsarticle}

\usepackage{amssymb}
\usepackage{amsmath}  	
\usepackage{url}

\newcounter{bla}

\renewcommand{\d}[2]{\frac{d #1}{d #2}} 
\renewcommand{\v}[1]{\ensuremath{\mathbf{#1}}} 

\usepackage [english]{babel}
\usepackage [english = american]{csquotes}
\MakeOuterQuote{"}

\journal{\null}

\begin{document}

\begin{frontmatter}



\title{Lagrange2D: A Mathematica package for Lagrangian analysis of two-dimensional fluid flows}


\author[a,b]{William Gilpin\corref{author}}

\cortext[author] {Corresponding author.\\\textit{E-mail address:} wgilpin@stanford.edu}
\address[a]{Department of Applied Physics, Stanford University}
\address[b]{Quantitative Biology Initiative, Harvard University}

\begin{abstract}

We introduce Lagrange2D, a Mathematica package for analysis and characterization of complex fluid flows using Lagrangian transport metrics. Lagrange2D includes built-in functions for integrating ensembles of trajectories subject to time-varying two-dimensional flows, as well as utilities for calculating various quantities of interest, such as finite-time Lyapunov exponents, stretching vector fields, the fractal dimension, and flushing times. The package also includes tools for visualizing transport and pathlines, as well as for generating videos. This package aims to ease rapid characterization of arbitrary flows, by allowing identification of Lagrangian coherent structures and other quantities of interest. The open-source code for the package is available on GitHub at: \url{https://github.com/williamgilpin/lagrange2d}

%

%

\end{abstract}


\end{frontmatter}



{\bf PROGRAM SUMMARY}

\begin{small}
\noindent
{\em Program Title:}    Lagrange2D                                      \\
{\em Licensing provisions:} GPLv3                                    \\
{\em Programming language:}  Mathematica 10.0 or higher                                  \\


{\em Nature of problem:} Time-varying flows are difficult to characterize using purely Eulerian measures, such as velocity and vorticity fields. We introduce Lagrange2D, a Mathematica package for efficiently simulating and analyzing two-dimensional, time-varying fluid flows, which also calculates and visualizes standard quantities associated with these flows, including Lyapunov exponents. Our package also contains utilities for generating videos of transport in time-varying flows, given a time-dependent velocity field as an input. \\
{\em Solution method:} Lagrange2D numerically integrates a set of trajectories originating from a given ensemble of initial conditions via a variable-step integration scheme. It then numerically interpolates a displacement map associated with changes in the geometry of the ensemble, and then computes finite-time Lyapunov exponents and other metrics using finite-differences operations on the interpolated displacement map. \\
{\em Additional comments including Restrictions and Unusual features:} Lagrange2D has no external dependencies, and consists primarily of self-contained, single functions that share a common call structure. Full documentation and a runnable "demos" notebook are included in the GitHub repository hosting the source code.   \\
   \\

\end{small}


The complex motion of fluids is traditionally analyzed using Eulerian metrics, such as the time-varying velocity field, as well as derived quantities such as the vorticity and shear fields. These quantities can fail to provide insight into complex, time-varying flows, which may advect fluid parcels along nonlinear and even chaotic trajectories \cite{bennett2006lagrangian,samelson2013lagrangian}. For such flows, analysis using a Lagrangian framework, which studies the evolution of the fluid in a comoving frame, can yield insight into the interplay between local dynamics and large-scale transport properties of the fluid \cite{price2006lagrangian}.

A limiting feature of Lagrangian analysis is its reliance on numerical integration of trajectories associated with the fluid in order to estimate relevant features, such as bulk flows and pair dispersion \cite{bennett2006lagrangian,price2006lagrangian}. This bounds the accuracy of the analysis to the maximum timescale for which trajectories can be stably integrated throughout the domain \cite{falkovich2011fluid}. However, recent conceptual and methodological advancements have allowed deeper insight into the global structure of unsteady flows to be obtained from even limited-duration Lagrangian simulations. In particular, finite-time Lyapunov exponents (FTLE), which can be calculated efficiently from trajectories originating from a mesh of initial conditions within the region of interest, can be used to identify kinetic barriers to transport in a flow \cite{haller2000lagrangian}. Ridges in the maximum FTLE field can be used to identify and demarcate "Lagrangian coherent structures," which are characteristic fluidic features (such as eddies) that form a "skeleton" of material barriers that dictate the transport properties of the flow \cite{haller2015lagrangian,shadden2011lagrangian,haller2000lagrangian}. Lagrangian coherent structures have been shown to strongly predict the long-term transport properties of diverse systems ranging from geophysical and atmospheric flows, to biological fluid flows \cite{haller2015lagrangian,gilpin2017flowtrace,shadden2006lagrangian}.

Several software packages exist for calculating FTLE fields using MATLAB and Python \cite{onu2015lcs,fredj2016particle,shadden2006lagrangian,peng2009transport}. However, for certain applications (such as theoretical analysis of analytically-defined flows) the ability to compute FTLE fields (and other, related, Lagrangian quantities) from within a symbolic algebra program may allow more straightforward characterization and comparison of unsteady flows. To this end, here we describe \texttt{Lagrange2D}, a Mathematica package that allows straightforward simulation of transport in time-varying velocity fields, as well as calculation of FTLE fields, flushing times, and other quantities of interest for two-dimensional flows. Our package also includes utilities for generating pathline visualizations and animations of two-dimensional flow fields. 

\section{Installation}

\texttt{Lagrange2D} requires Mathematica 10.0 or higher. The source code is available on GitHub: \texttt{https://github.com/williamgilpin/lagrange2d}. After downloading the repository, the functions contained in the package can be accessed in a local notebook by running the following (with an appropriate absolute path substituted):

\begin{verbatim}
Import["/Users/me/path/to/folder/lagrange2d/lagrange2d.wl"]
\end{verbatim}

Alternatively, the file \texttt{lagrange2d.wl} can be manually moved into the working directory of the local notebook, in which case the following lines will import the package:

\begin{verbatim}
SetDirectory[NotebookDirectory[]]
<< lagrange2d.wl
\end{verbatim}

After successful installation, detailed and executable examples demonstrating the various functions in the package are available in the included \texttt{demos.nb} notebook.

\section{Included functions}

We summarize in Table \ref{funcs} the different functions included in Lagrange2D. Here, we highlight several functions and their underlying theoretical basis.

\begin{table}[htp]
\centering
\caption{Functions included in Lagrange2D. Full documentation and runnable examples are included in the notebook "demos.nb" on GitHub at \texttt{https://github.com/williamgilpin/lagrange2d/blob/master/demos.nb} }
\vspace{4 mm}
\label{funcs}
\begin{tabular} {|p{5cm}|p{6cm}|}
\hline
Function & Description\\
\hline
\texttt{advectPoints} & Numerically advect of a set of initial conditions\\
\hline
\texttt{findFTLEField}		& Find all finite-time Lyapunov exponents\\
\hline
\texttt{findMaxFTLEField}		& Find the maximum finite-time Lyapunov exponents\\
\hline
\texttt{findStretchlines}		& Find the vector fields corresponding to maximum and minimal stretching\\
\hline
\texttt{findKYDim}	& Find the Kaplan-Yorke fractal dimension\\
\hline
\texttt{flushingTimes}		& Find the flushing time\\
\hline
\texttt{pathPlot}		& Draw pathlines of fixed duration for a set of initial conditions\\
\hline
\texttt{animateFlow}		& Generate a movie file showing the temporal evolution of a set of initial conditions\\
\hline
\texttt{makeMesh} 	& Draw a uniform rectangular mesh\\
\hline
\end{tabular}
\end{table}

\subsection{Finite-time Lyapunov Exponents}

The Lyapunov exponents of a dynamical system quantify the degree to which trajectories originating at given points in the domain spread apart over time. While these formally may be calculated by computing trajectories of pairs of neighboring points on the attractor, recent work has shown that they are well-approximated by the finite-time Lyapunov exponent \cite{haller2015lagrangian,shadden2011lagrangian}, defined as
\begin{equation}
\lambda(\v r) \equiv \dfrac{1}{\tau}\max\left[\text{eig}\left(			\log\left(	\d{\phi_{0}^{\tau}(\v r)}{\v r}	\right)		\right)\right]
\end{equation}
where the flow map $\phi_{0}^{\tau}$ transforms a tracer particle at initial position $\v r$ to its final position $\v r'$  after a time $\tau$. In practice, this requires numerically integrating an ensemble of trajectories for a duration $\tau$, and then numerically interpolating the map $\phi_{0}^{\tau}$.

In the functions \texttt{findFTLEField} and \texttt{findMaxFTLEField}, a region of initial conditions is specified, and then advected forward in time for a specified duration. The displacement map $\phi_{0}^{\tau}$ is calculated using Mathematica's built-in \texttt{Interpolate} function.

\subsection{Kaplan-Yorke Fractal Dimension}

The Kaplan-Yorke fractal dimension quantifies the structure of a dynamical system's attractor \cite{kaplan1979functional}. Unlike the box-counting and information dimensions, the Kaplan-Yorke dimension has the advantage of being directly computable from particle trajectories, making it well-suited for the analysis of unsteady flows. For a general dynamical system, the Kaplan-Yorke dimension is defined as
\begin{equation}
D_{KY} \equiv j+{\frac  {\sum _{{i=1}}^{j}\lambda _{i}}{|\lambda _{{j+1}}|}},
\label{ky}
\end{equation}
where the Lyapunov exponents of the system, $\lambda_i$, have been ranked from largest to smallest, and $j$ denotes the first index that satisfies the relationship
\[
\sum _{i=1}^{j}\lambda _{i}\geqslant 0, \qquad \sum _{{i=1}}^{{j+1}}\lambda _{i}<0.
\]
In two dimensional, incompressible flows, $j = 1$. However, when using estimates of $\lambda$, such as those inferred from experimental velocity fields, or those resulting from differential methods like particle image velocimetry, strict incompressibility is not necessarily preserved, and so $j \in \{0,1,2\}$. In the function \texttt{findKYDim} in \texttt{Lagrange2D}, we use estimates of the Lyapunov exponents generated by the FTLE calculation in \eqref{ky}, in order to produce a timescale-dependent scalar field $\mathcal D_{KY}^t(x,y)$.

In fluid flows, the Kaplan-Yorke dimension quantifies the tendency of particles to cluster within subregions of the fluid's attractor: wherever the Kaplan-Yorke dimension equals the dimensionality of the dependent variables ($d=2$ here), then there is no particular tendency of trajectories to localize \cite{vilela2017map,tuval2004opening}. In contrast, in subregions where the Kaplan-Yorke dimension is much less than the dimension of the underlying dynamics, particles will tend to cluster and form filamentary structures. In periodic or statistically stationary flows, the spatially-resolved Kaplan-Yorke dimension can therefore be used to differentiate the stable and unstable manifolds of the flow.

\section{Example Usage}

\begin{figure}
{
\centering
\includegraphics[width=\linewidth]{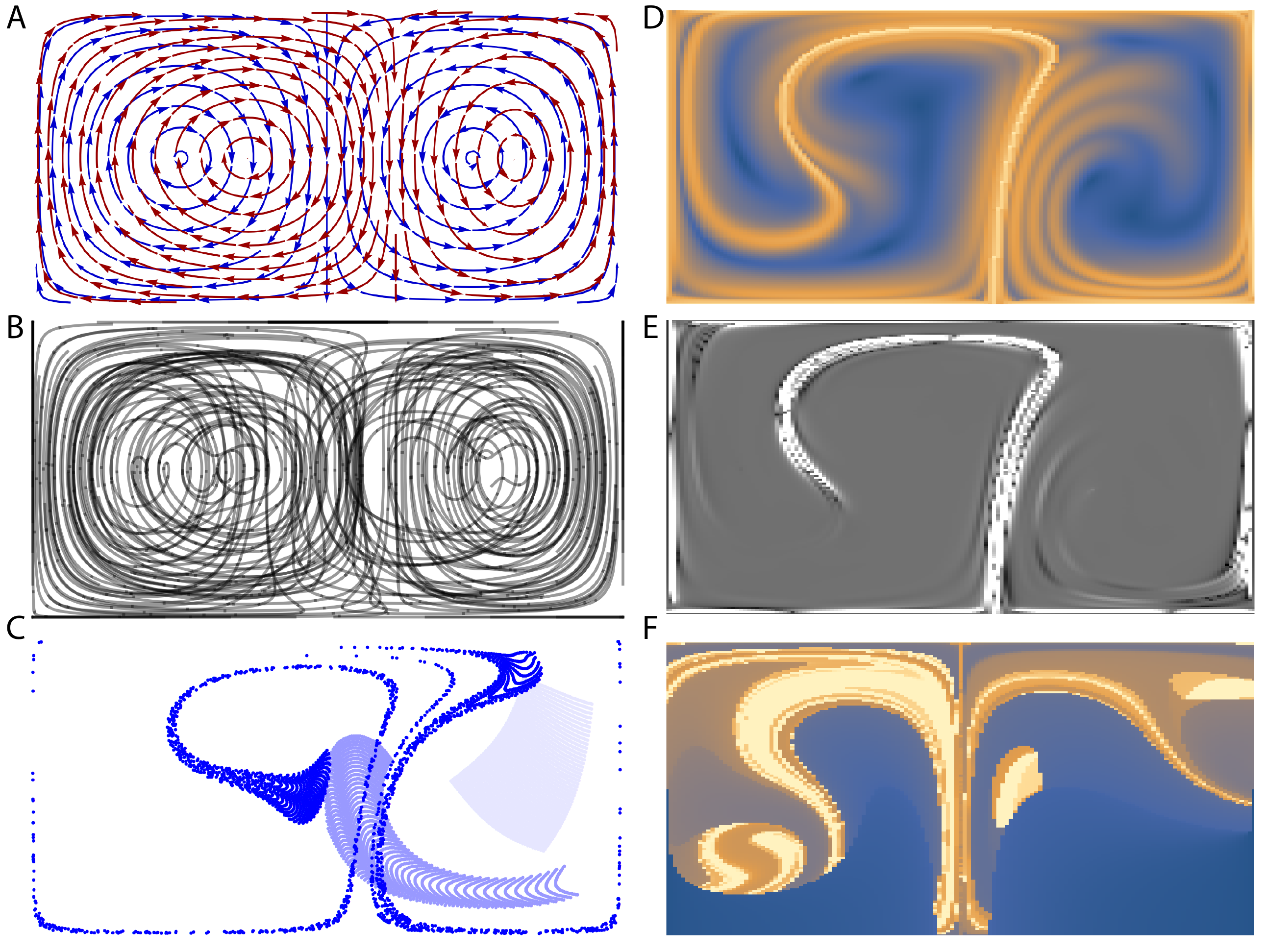}
\caption{
{\bf Example plots produced by Lagrange2D.} (A) Streamlines of the time-dependent "double gyre" flow, generated at times corresponding to times when the flow is maximally symmetric (blue) and maximally asymmetric (red). (B) Pathlines for particles advected by the double gyre flow, generated using the function \texttt{pathPlot}. (C) The motion of a blob of particles advected by the flow, with successive timepoints indicated in darkening shades of blue. Plot generated using the function \texttt{advectPoints}. (D) The maximal finite-time Lyapunov exponents, generated using the function \texttt{findFTLE}. (E) The Kaplan-Yorke fractal dimension, generated using the function \texttt{findKYDim}. (F) The flushing times for the shown domain, generated using the function \texttt{flushingTimes}. For these images, the double gyre flow was used, with parameters $A=0.1,\epsilon=0.25,\omega=\pi/5$.
}
\label{fig_lagrange}
}
\end{figure}

The following example and images are available in executable form in the \texttt{demos.nb} notebook included in the installation directory. Plots produced by the examples below are shown in Figure \ref{fig_lagrange}, and full documentation of arguments and keyword arguments for each function can be accessed via the function's docstring. For example, to access the documentation for the function \texttt{pathPlot}, the command \texttt{?pathPlot} can be executed anywhere within an active notebook.

\subsubsection{Define a velocity field}
Figure \ref{fig_lagrange} shows applications of the functions described in the previous section to the study of the double gyre flow, which has the form
\begin{align*}
v_x(x,y,t) &= - \pi A \sin(\pi f(x,t)) \cos(\pi y), 			\\
v_y(x,y,t) &= \pi A \cos(\pi f(x,t)) \sin(\pi y) \big(2 a(t) x + b(t)\big)
\end{align*}
where
\begin{align*}
a(t) &= \epsilon \sin(\omega t), \\
b(t) &= 1-2\epsilon \sin(\omega t), \\
f(x,t) &= a(t) x^2 + b(t) x.
\end{align*}
The parameters $\omega$ and $A$ control the transport timescales of the flow, while $\epsilon$ determines the degree of temporal variation, and thus chaotic mixing present \cite{shadden2005definition}.

When using \texttt{Lagrange2D}, this flow may be defined as a standard Mathematica function,
\begin{verbatim}
a[t_] := e Sin[w t];
b[t_] := 1 - 2 e Sin[w t];
f[t_, x_] := a[t] x^2 + b[t] x;

g[t_, x_, y_] := Pi A Sin[Pi f[t, x]] Cos[Pi f[t, y]](*D[f[t,x],x]*)

(* double gyre flow *)
{vx[t_, x_, y_],  vy[t_, x_,  y_]} := 
     {-Pi A Sin[Pi f[t, x]] Cos[Pi y], 
       Pi A Cos[Pi f[t, x]] Sin[Pi y] (2 a[t] x + b[t])};

params = {A -> 0.1, e -> 0.1, w -> Pi/5};
\end{verbatim}

\subsection{Visualize and simulate a time-varying flow}

Plotting the pathlines associated with a flow uses similar syntax to Mathematica's built-in \texttt{StreamPlot} function,
\begin{verbatim}
pathPlot[{vx[t, x, y], vy[t, x, y]} /. params, 
  {x, 0, 2}, {y, 0, 1}, {t, 0, 15}]
\end{verbatim}
The density and spacing of pathlines can be manipulated using similar optional arguments to those used by \texttt{StreamPlot}, which are further described in the function's docstring.

Similar syntax is used to integrate the motion of a set of initial conditions subject to the flow. However, the full initial conditions must be defined by the user in a separate command,
\begin{verbatim}
blobPoints = makeMesh[{.2, .6}, {.1, .5}, 50, 50];
ptLocs = advectPoints[ 
   blobPoints,
   {vx[t, x, y], vy[t, x, y]} /. params,
   {t, 0, 20}, x, y];
\end{verbatim}

The motion of this collection of points may be visualized by exporting a video,
\begin{verbatim}
animateFlow[ 
  {vx[t, x, y], vy[t, x, y]} /. params,
  blobPoints, {t, 0, 20},{x, 0, 2}, {y, 0, 1}, 20, 1/10];
\end{verbatim}

\subsection{Find the maximal FTLE field}

Finding the FTLE field associated with a velocity field uses similar syntax to the commands described above. This function requires the exact mesh points at which to compute the FTLE field to be specified by the user.
\begin{verbatim}
domainPoints = makeMesh[{0, 2}, {0, 1}, 150, 150];
ftleField = 
  findMaxFTLEField[{vx[t, x, y], vy[t, x, y]} /. params, 
   domainPoints, {t, 0, 10}, x, y];
\end{verbatim}

\subsection{Find a Kaplan-Yorke fractal dimension field}

In order to calculate the Kaplan-Yorke fractal dimension, two commands are required: one to generate the finite-time Lyapunov exponents associated with each point in a user-specified mesh, and another to convert these exponents into an estimate of the Kaplan-Yorke fractal dimension,
\begin{verbatim}
domainPts = makeMesh[{0, 2}, {0, 1}, 150, 150];
ftleVals = 
  findFTLEField[{vx[t, x, y], vy[t, x, y]} /. params, 
   domainPoints, {t, 0, 10}, x, y];
kyVals = findKYDim[ftleVals];
\end{verbatim}

\subsection{Find the stretching vector field}

The stretching distribution corresponds to a vector field defined at each point in the domain. Similar to the above commands for computing FTLE fields, this function requires a mesh to be specified by the user. There are two orthogonal stretching fields, one associated with the rate of maximal stretching, and the other associated with the rate of minimal stretching,
\begin{verbatim}
blobPoints = makeMesh[{0, 2}, {0, 1}, 100, 100];
stretchVecsPositive = 
  findStretchlines[{vx[t, x, y], vy[t, x, y]} /. params, 
   blobPoints, {t, 0, 10}, x, y, "Positive"];
stretchVecsNegative = 
  findStretchlines[{vx[t, x, y], vy[t, x, y]} /. params, 
   blobPoints, {t, 0, 10}, x, y, "Negative"];
\end{verbatim}

\subsection{Find the flushing time field}

The flushing, or residence, time distribution has similar syntax to Mathematica's built-in \texttt{StreamPlot} function. However, because the boundaries of the domain determine the residence time, this field depends directly on the bounds of the specified mesh.
\begin{verbatim}
flushField = flushingTimes[
   {vx[t, x, y], vy[t, x, y]} /. params,
   {x, -1, 1},{y, 0, 1},{t, 0, 50},{n -> 20000}];
\end{verbatim}

The results of some of the functions described above are showing in Figure \ref{fig_lagrange}. Comparison of snapshots of advected particle positions, and the FTLE field, reveals that the FTLE field successfully reveals kinetic barriers to transport in the flow. Comparison of the FTLE field and advection of a group of particles shows that groups particles tend to deform around regions of high FTLE values. The Kaplan-Yorke dimension indicates regions of the flow that are kinetically accessible within the integration time, and it establishes regions near which trajectories tend to cluster. Consistent with the pathlines of the flow, the flushing time plot reveals asymmetry intrinsic to the map, indicated by strong differences in the residence time of particles depending on whether they originate in regions to the left or right of the unstable manifold of the flow.

\section{Conclusions}

We present a Mathematica package, \texttt{Lagrange2D}, for the analysis and visualization of transport in two-dimensional fluid flows. Our package has particular relevance to characterizing unsteady flows, for which standard velocity fields and streamline visualization can fail to capture the full structure of the flow. \texttt{Lagrange2D} includes utilities for integrating ensembles of trajectories, as well as functions for computing finite-time Lyapunov exponents, the fractal dimension, and the flushing time distribution. Additionally, our package contains visualization tools for viewing pathlines and for generating videos of time-varying flows. \texttt{Lagrange2D} is self-contained and simple to install, and we hope that it will simplify the characterization of two-dimensional flows, particularly those for which the velocity field has an analytic form.

\section{Acknowledgements}

W.G. thanks the U. S. Department of Defense for their support through the NDSEG Fellowship program. The author declares no competing interests.





\bibliographystyle{elsarticle-num}
\bibliography{lagrange_cites.bib}







\end{document}